\documentstyle{mn}

\newif\ifAMStwofonts

\ifoldfss/home
  \ifCUPmtlplainloaded \else
    \NewTextAlphabet{textbfit} {cmbxti10} {}
    \NewTextAlphabet{textbfss} {cmssbx10} {}
    \NewMathAlphabet{mathbfit} {cmbxti10} {} 
    \NewMathAlphabet{mathbfss} {cmssbx10} {} 
  \fi
  \ifAMStwofonts
    \ifCUPmtlplainloaded \else
      \NewSymbolFont{upmath} {eurm10}
      \NewSymbolFont{AMSa} {msam10}
      \NewMathSymbol{\upi}     {0}{upmath}{19}
      \NewMathSymbol{\umu}     {0}{upmath}{16}
      \NewMathSymbol{\upartial}{0}{upmath}{40}
      \NewMathSymbol{\leqslant}{3}{AMSa}{36}
      \NewMathSymbol{\geqslant}{3}{AMSa}{3E}

      \let\leq=\leqslant 
       
    \fi
  \fi
\fi 

\ifnfssone
  \newmathalphabet{\mathit}
  \addtoversion{normal}{\mathit}{cmr}{m}{it}
  \addtoversion{bold}{\mathit}{cmr}{bx}{it}
  \newmathalphabet{\mathbfit} 
  \addtoversion{normal}{\mathbfit}{cmr}{bx}{it}
  \addtoversion{bold}{\mathbfit}{cmr}{bx}{it}
  \newmathalphabet{\mathbfss} 
  \addtoversion{normal}{\mathbfss}{cmss}{bx}{n}
  \addtoversion{bold}{\mathbfss}{cmss}{bx}{n}
  \ifAMStwofonts
    \ifCUPmtlplainloaded \else
      \UseAMStwoboldmath
      \makeatletter
      \new@mathgroup\upmath@group
      \define@mathgroup\mv@normal\upmath@group{eur}{m}{n}
      \define@mathgroup\mv@bold\upmath@group{eur}{b}{n}
      \edef\UPM{\hexnumber\upmath@group}
      \new@mathgroup\amsa@group
      \define@mathgroup\mv@normal\amsa@group{msa}{m}{n}
      \define@mathgroup\mv@bold\amsa@group{msa}{m}{n}
      \edef\AMSa{\hexnumber\amsa@group}
      \makeatother
      \mathchardef\upi="0\UPM19
      \mathchardef\umu="0\UPM16
      \mathchardef\upartial="0\UPM40
      \mathchardef\leqslant="3\AMSa36
      \mathchardef\geqslant="3\AMSa3E

      \let\leq=\leqslant 

    \fi
  \fi
\fi 

\ifnfsstwo
  \DeclareMathAlphabet{\mathbfit}{OT1}{cmr}{bx}{it}
  \SetMathAlphabet\mathbfit{bold}{OT1}{cmr}{bx}{it}
  \DeclareMathAlphabet{\mathbfss}{OT1}{cmss}{bx}{n}
  \SetMathAlphabet\mathbfss{bold}{OT1}{cmss}{bx}{n}
  \ifAMStwofonts
    \ifCUPmtlplainloaded \else
      \DeclareSymbolFont{UPM}{U}{eur}{m}{n}
      \SetSymbolFont{UPM}{bold}{U}{eur}{b}{n}
      \DeclareSymbolFont{AMSa}{U}{msa}{m}{n}
      \DeclareMathSymbol{\upi}{0}{UPM}{"19}
      \DeclareMathSymbol{\umu}{0}{UPM}{"16}
      \DeclareMathSymbol{\upartial}{0}{UPM}{"40}
      \DeclareMathSymbol{\leqslant}{3}{AMSa}{"36}
      \DeclareMathSymbol{\geqslant}{3}{AMSa}{"3E}

      \let\leq=\leqslant 

    \fi
  \fi
\fi 

\ifCUPmtlplainloaded \else
  \ifAMStwofonts \else 
    \def\upi{\pi}
    \def\umu{\mu}
    \def\upartial{\partial}
  \fi
\fi

\title[Global VLBI Observations of M82]
  {Global VLBI Observations of Compact Radio Sources in M82}
\author[A.~R.~McDonald et al.]
  {A.~R.~McDonald,$^1$\thanks{email: amd@jb.man.ac.uk} T.~W.~B.~Muxlow,$^1$ A.~Pedlar,$^1$ M.~A.~Garrett,$^2$\newauthor K.~A.~Wills,$^{3}$ S.~T.~Garrington,$^1$ P.~J.~Diamond,$^1$ P.~N.~Wilkinson$^1$
  \\$^1$University of Manchester, Jodrell Bank Observatory, Macclesfield, Cheshire SK11 9DL.\\
  $^2$Joint Institute for VLBI in Europe, Postbus 2, 7990 Dwingeloo, The Netherlands\\
  $^3$Department of Physics, University of Sheffield, Sheffield S3 7RH.}
\date{Accepted ?. Received ?}

\pagerange{\pageref{firstpage}--\pageref{lastpage}}
\pubyear{2000}

\begin{document}

\maketitle

\label{firstpage}
\begin{abstract}
Observations of the starburst galaxy, M82, have been made with a 20-station global VLBI array at $\lambda$18cm. Maps are presented of the brightest young supernova remnants (SNR) in M82 and the wide-field mapping techniques used in making images over a field of view of $\sim$1 arcminute with 3 milliarcsecond resolution are discussed. A limit has been placed on the power law deceleration of the young SNR, 43.31+592 with an index greater than 0.73 $\pm$ 0.11 from observations with the European VLBI Network. Using the global array we have resolved compact knots of radio emission in the source which, with future global observations, will enable better constraints to be placed on the expansion parameters of this SNR.

The latest global observations have also provided high resolution images of the most compact radio source in M82, 41.95+575. We determine an upper limit to the radial expansion rate along the major axis of 2000 km s$^{-1}$. However, the new images also show structure resembling that of collimated ejection which brings into question the previous explanation of the source as being due to the confinement of a supernova by a high density circumstellar medium.

It is apparent that we are now able to image the brightest supernova remnants in M82 with a linear scale which allows direct comparison with galactic SNR such as Cassiopeia A.
\end{abstract}
\section{Introduction}
M82 is generally considered to be the archetypal `starburst' galaxy. Far-infrared (FIR) and radio luminosities imply a high star-formation rate and at a distance of 3.2 Mpc \cite{burbidge64} it lends itself well to study with radio interferometry on a parsec scale. The radio studies of M82 have revealed a population of compact radio sources (Unger et al., 1984; Kronberg et al., 1985)\nocite{unger84}\nocite{kronberg85} which were tentatively identified as being either radio supernovae (RSn) or supernova remnants (SNR). Subsequent observations with MERLIN at a resolution of 50 mas (0.75 pc) resolved most of the compact sources and the parsec-scale shell-like structure observed in many of the objects confirmed them to be young SNR \cite{muxlow94}. Furthermore, recent EVN observations at 15 mas (0.2 pc) resolution \cite{pedlar99} have further resolved the most compact sources and by comparison with a 1986 epoch an expansion speed of $\sim$9500 kms$^{-1}$ has been measured for the compact shell-like remnant, 43.31+592. Obviously it would be preferable to measure the size evolution of this remnant on much shorter timescales than the $\sim$10 years required by the EVN observations. A global VLBI array at $\lambda$18cm provides a maximum resolution of $\sim$3 mas and enables these measurements to be performed on timescales of $\sim$2 years and also provides an opportunity to determine the deceleration of the remnants as they interact with the interstellar medium (ISM). The SNR in M82 and other nearby galaxies provide a means of understanding the star-formation processes in a starburst environment and also enable the ISM in a starburst to be probed on parsec scales. 

Since the SNR are produced by massive stars with lifetimes of order 10$^{6}$ years, then these SNR trace the recent star formation history of a galaxy. If the supernova rate can be established, then by assuming an initial mass function it is possible to calculate a total star formation rate to compare with that derived from other methods, such as the total far-infrared or non-thermal radio luminosity. Previous VLBI observations of M82 have concentrated on the brightest and most compact source, 41.95+575 (Bartel et al., 1987; Trotman 1996)\nocite{trotman96}\nocite{bartel87}. However, we are now able to process the large amounts of data needed to image a large field of view. Therefore, it is now possible to image a field of view of $\sim$1 arcminute with the 3 mas resolution which the global array provides at $\lambda$18cm.
\section{The observations and image processing}
M82 and a nearby phase calibrator, 0955+697, were observed at 1652 MHz with 20 antennas of a global VLBI network on 28th November, 1998. The network comprised all 10 antennas of the Very Long Baseline Array (VLBA), 7 antennas of the European VLBI Network (EVN), the two NASA Deep Space Network (DSN) antennas at Robledo, Spain and Goldstone, USA and a single VLA dish. The telescopes received both left and right hand circular polarisations (LCP \& RCP) which, after correlation, were combined to form the total intensity maps presented in this paper. The observations were made with the VLBA/MkIV recording system and a total bandwidth of 16 MHz. The observations were carried out in phase reference mode with a cycle of 8 minutes on the target (M82) and 3 minutes on the phase calibrator. At the time of correlation, this was the largest VLBI experiment ever performed and the 190 baselines provided by the combinations of 20 stations resulted in excellent UV-coverage which in turn provided a greater image fidelity than had previously been achieved.

The data were correlated at the National Radio Astronomy Observatory, Socorro, New Mexico, using the VLBA processor. The correlator used a pre-averaging time of 2 seconds and a frequency resolution of 0.25 MHz. Three separate correlations of the data were performed in order to help provide the $\sim$1 arcminute field of view required to image the SNR. Each correlation was centred on the position of a bright compact source and these positions are given in Table 1. 
\begin{table}
\begin{center}
\caption{Positions used for the three correlator runs on the global VLBI data.}
\begin{tabular}{|c|c|c|c|}\hline
& Source & Right Ascension & Declination \\ 
& & (J2000) & (J2000)\\ \hline
1 & 41.95+575 & 09$^{h}$ 55$^{m}$ 50.694$^{s}$ & 69$^{\circ}$ 40' 43.69'' \\ 
2 & 43.31+592 & 09$^{h}$ 55$^{m}$ 52.030$^{s}$ & 69$^{\circ}$ 40' 45.42'' \\
3 & 44.01+596 & 09$^{h}$ 55$^{m}$ 52.720$^{s}$ & 69$^{\circ}$ 40' 45.78'' \\ \hline
\end{tabular}
\label{correlatetable}
\end{center}
\end{table}

All subsequent processing was carried out using the NRAO AIPS package. The visibility amplitudes were calibrated using the system temperature and gain information provided for each telescope and the residual delays, fringe rates and antenna gains were determined from 0955+697 and interpolated and applied to M82. In order to avoid time-averaging and bandwidth smearing, the data were kept in the form of 64 continuous 0.25 MHz channels and were not averaged in time. Initially, the data correlated at the position of the brightest source, 41.95+575, were Fourier transformed and CLEANed using the AIPS task IMAGR. The solar activity at the time of observation was high so that the phase stability during the observing run was not good. Therefore, the phase solutions derived from the nearby phase calibrator, 0955+697, were not sufficient to completely remove atmospheric phase fluctuations in the direction of M82. Therefore, further improvements were made using standard self-calibration techniques. The delay, fringe rate and phase solutions derived from this initial fringe-fitting and self-calibration procedure were then applied to the other two pointings which were then Fourier transformed and CLEANed. The increase in resolution from 15 mas with the EVN \cite{pedlar99} to 3 mas results in a decrease in surface brightness sensitivity. Hence, only two of the five sources detected with the EVN are detected at full resolution. However, by applying a Gaussian taper to the data in the UV-plane, maps were made at 10 mas and 15 mas resolution for comparison with the EVN images. 

\begin{figure*}
\begin{center}
\setlength{\unitlength}{1cm}
\begin{picture}(5,6)(0,0)
\put(-6,0){\includegraphics{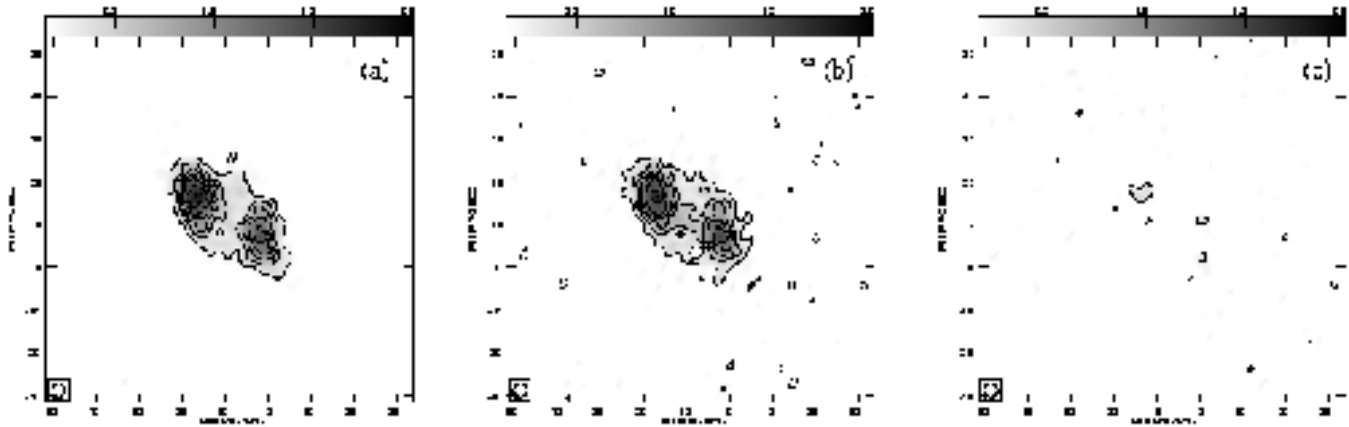}}
\end{picture}
\caption{18cm global VLBI images of the compact source 41.95+575 in M82 at 3 mas resolution. (a) was made from data correlated to the position of 41.95+575, (b) was made from data correlated to the position of 43.31+592 and (c) is the difference between (a) and (b). The greyscale ranges from 0.2 to 2 mJy beam$^{-1}$ and the contour levels are (-1, 1, 2,..., 9, 10) $\times$ 0.2 mJy beam$^{-1}$.}
\label{4195montage}
\end{center}
\end{figure*}

\begin{figure*}
\begin{center}
\setlength{\unitlength}{1cm}
\begin{picture}(5,6)(0,0)
\put(-6,0){\includegraphics{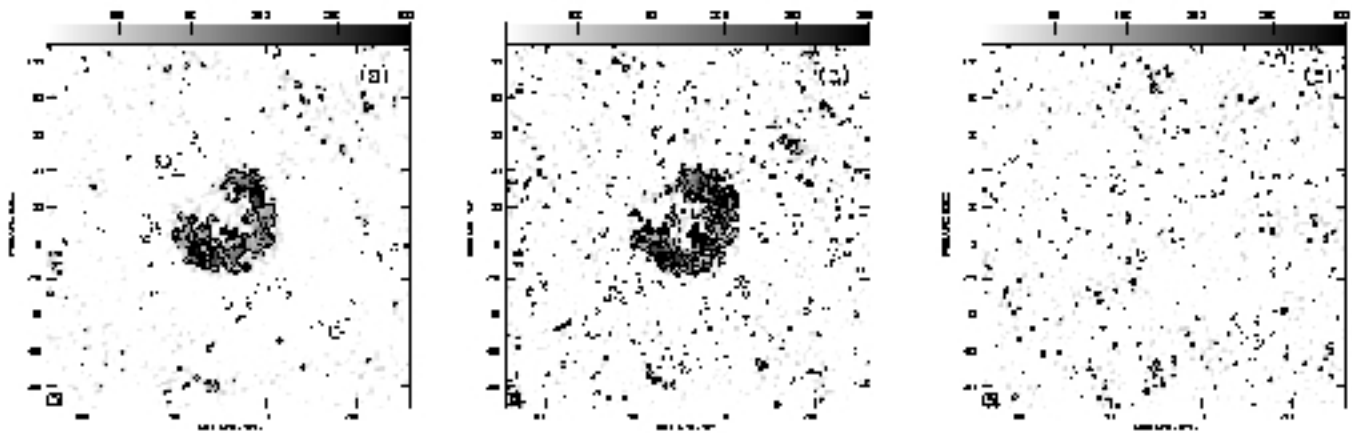}}
\end{picture}
\caption{18cm global VLBI images of the young supernova remnant, 43.31+592 in M82 at 4 mas resolution. (a) was made from data correlated to the position of 41.95+575, (b) was made from data correlated to the position of 43.31+592 and (c) is the difference between (a) and (b). The greyscale ranges from 0.05 to 0.3 mJy beam$^{-1}$ and the contour levels are (-1, 1, 1.5, 2, 2.5, 3) $\times$ 0.1 mJy beam$^{-1}$.}
\label{4331montage}
\end{center}
\end{figure*}

The 3 mas resolution maps made from the data correlated at the position of the brightest source, 41.95+575, had a rms noise off-source of 45 $\mu$Jy beam$^{-1}$. However, the same phase and amplitude corrections when applied to the other two correlations were only sufficient to produce maps with noise levels of 68 $\mu$Jy beam$^{-1}$ at the same resolution. This difference in sensitivity between different correlations illustrates the importance of having a bright source with which self-calibration can be performed near to the pointing centre of the observations. Therefore, maps made from the data correlated at the positions of 43.31+592 and 44.01+596 were only used as a consistency check. However, the ability to make wide-field maps and hence image 43.31+592 from data correlated at the position of 41.95+575 emphasises the power of correlating the data with a wide-field mode (i.e. short integration times and small channel bandwidths). A comparison between the 3 mas resolution maps of 41.95+575 made from pointings 1 and 2 is shown in figure~\ref{4195montage}. To check for consistency at slightly lower resolution, figure~\ref{4331montage} shows a comparison between the maps of 43.31+592 also made from pointings 1 and 2 and restored with a CLEAN beam of FWHM 4 mas.

\begin{figure*}
\begin{center}
\setlength{\unitlength}{1cm}
\begin{picture}(0,0)(0,0)
\put(0,0){\includegraphics{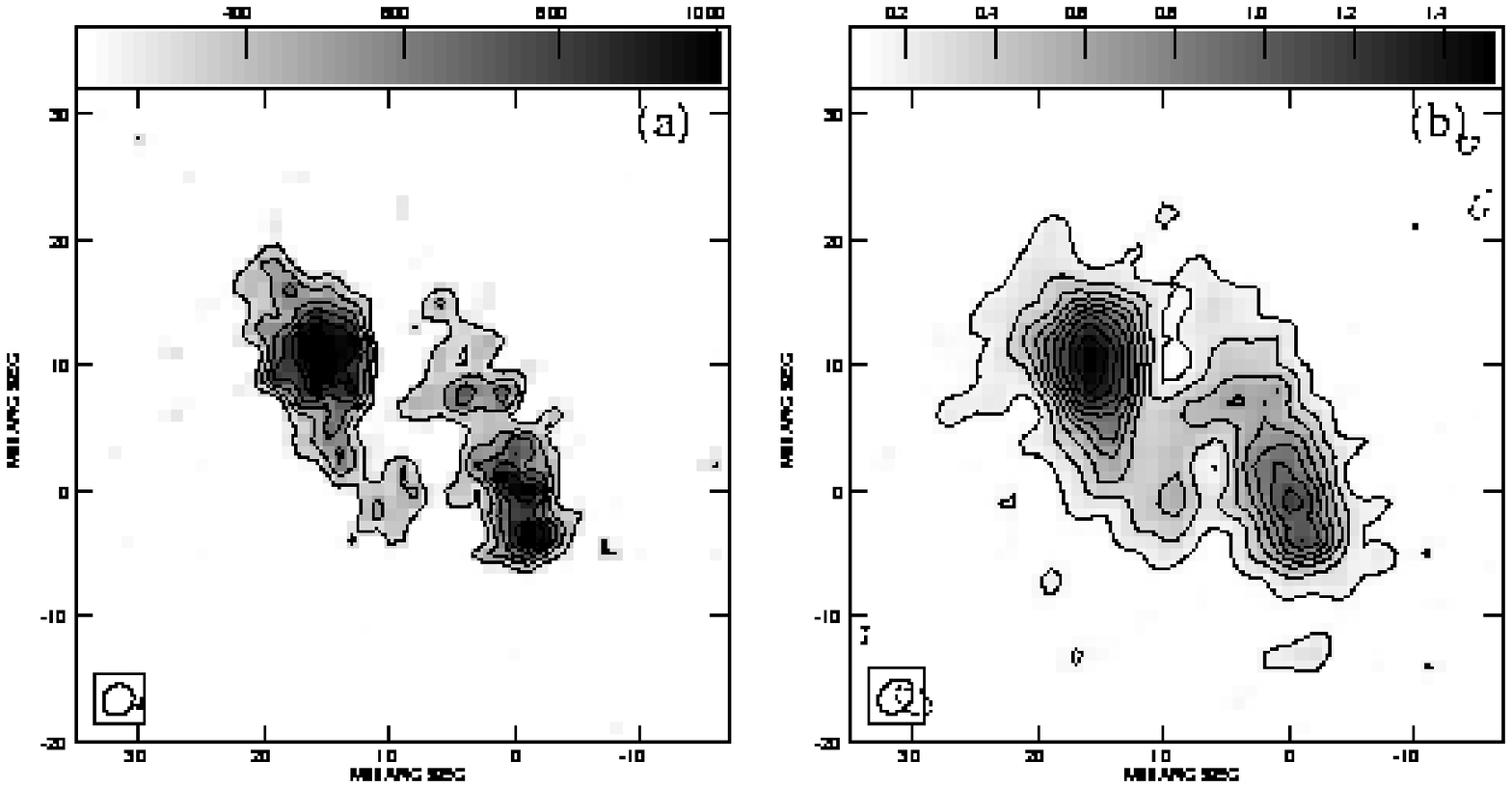}}
\end{picture}
\vspace{8cm}
\caption{Images of the most compact radio source in M82, 41.95+575. (a) was made from uniformly weighted data with Brigg's robustness set to -2 and restored with a 2.7 $\times$ 2.3 mas elliptical beam at a position angle of -45$^{\circ}$. The greyscale ranges from 0.2 to 1 mJy beam$^{-1}$ and the contour levels are (-1, 1, 1.5, 2, 2.5, 3, 3.5, 4) $\times$ 0.3 mJy beam$^{-1}$. (b) was made using an intermediate weighting scheme (robustness = 0) and restored with a 3.1 $\times$ 2.4 mas elliptical beam at a position angle of -44$^{\circ}$. The greyscale ranges from 0.1 to 1.5 mJy beam$^{-1}$ and the contour levels are (-1, 1, 2,..., 9, 10) $\times$ 0.15 mJy beam$^{-1}$.}
\label{4195final}
\end{center}
\end{figure*}

By applying a purely uniform weighting scheme (with the Brigg's robustness parameter \cite{briggs95} set to -2) to the UV data, a maximum resolution of $\sim$2.5 mas was achieved with a corresponding decrease in sensitivity to 88 $\mu$Jy beam$^{-1}$. However, of all the sources in M82, only 41.95+575 has a sufficient brightness to be detectable at this resolution and sensitivity and the resulting map is shown in figure~\ref{4195final}(a). In addition, figure~\ref{4195final}(b) shows the same source at slightly lower resolution.

\begin{figure*}
\begin{center}
\setlength{\unitlength}{1cm}
\begin{picture}(0,0)(0,0)
\put(0,0){\includegraphics{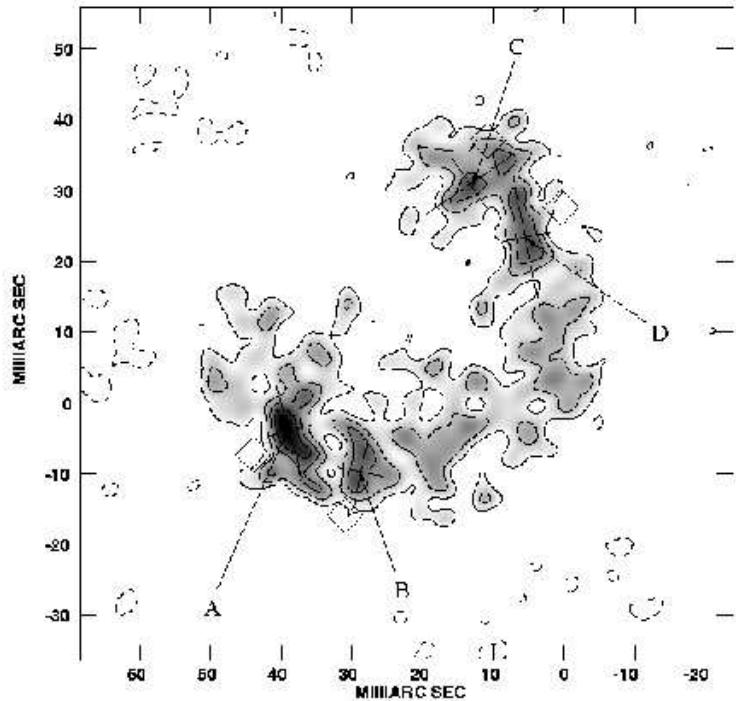}}
\end{picture}
\vspace{10cm}
\caption{Image of 43.31+592 made with uniformly weighted data (robustness = 0) and restored with a circular 4 $\times$ 4 mas$^{2}$ beam. The greyscale ranges from 0.1 to 0.3 mJy beam$^{-1}$ and the contour levels are (-1, 1, 1.5, 2, 2.5, 3) $\times$ 0.1 mJy beam$^{-1}$. Labels A to D indicate the four most significant knots of radio emission and the cross in the centre marks the average position of these knots. The diamond shapes represent the expected positions of these knots 10 years after the 1998.9 epoch assuming a constant self-similar radial expansion speed of 10000 km s$^{-1}$.}
\label{4331final}
\end{center}
\end{figure*}

A weighting scheme intermediate between the uniform weighting required to image 41.95+575 and pure natural weighting was applied to the data and images with a sensitivity sufficient to image the shell-like remnant, 43.31+592 were made.  Figure~\ref{4331final} shows the image of 43.31+592 at a resolution of 4 mas.

\section{Discussion}
\subsection{The nature of the compact source, 41.95+575}
\subsubsection{Radio morphology}
The brightest, most compact source in M82, 41.95+575 has been extensively studied with VLBI for around fifteen years. Observations at 2.3 GHz \cite{bartel87} and 5 GHz \cite{wilkinson90} showed structure which was interpreted to be an elongated shell and accounting for the fact that 41.95+575 was detected at epoch 1965.6 \cite{bash68}, limits were placed on its expansion speed of $\leq$ 14000 km s$^{-1}$. An extensive study of the source has been undertaken at $\lambda$18cm by Trotman (1996)\nocite{trotman96} and by comparing global VLBI images at epochs 1987.9 and 1990.7 a tentative expansion speed between the two `hotspots' of 2500 $\pm$ 550 km s$^{-1}$ was measured.

The new images of 41.95+575 show significant structure on smaller scales than have previously been observed with VLBI. In addition to the well established separation between the north-eastern and south-western `hotspots', significant structure is now observed linking the two hotspots as shown in figure~\ref{4195final}. The most compact structure shown in figure~\ref{4195final} was suggested in the 5 GHz EVN image of Wilkinson \& de Bruyn (1990) and has now been confirmed. From the 5 GHz map of Wilkinson \& de Bruyn (1990) we estimate a separation between the brightest components of 19 $\pm$ 1 mas. This compares with a similar estimate from the latest maps of 22 $\pm$ 1 mas. Therefore in the 13.1 years between these epochs, the source has expanded by 0.05 $\pm$ 0.02 pc (using a distance to M82 of 3.2 Mpc). However, given the uncertainties in identifying equivalent source structure between epochs and frequency bands, we prefer to express this as an upper limit to the expansion rate of 4000 km s$^{-1}$ (a radial expansion rate of 2000 km s$^{-1}$).

Various models have been proposed to account for the unusual morphology of 41.95+575.  Blondin et al. (1996)\nocite{blondin96} suggested that an axisymmetric density distribution in the wind from the progenitor star can lead to a protrusion emerging along the symmetry axis. They cite 41.95+575 as an example of this based on the VLBI images of Bartel (1997) and Wilkinson \& de Bruyn (1990). However, other models which predict nonspherical morphologies include that of Khokhlov et al. (1999)\nocite{khokhlov99} which involves the triggering of a supernova by supersonic jets at the centre of the progenitor star. The high resolution of the new global VLBI image reveals structure which may be interpreted as collimated outflow and we speculate that this may represent a jet-like structure such as that generated by the model of Khokhlov et al. (1999). However, it is not yet clear how this model could explain the high degree of confinement which the source appears to be experiencing.

\subsubsection{Flux density variability}
Of all the compact sources in M82, 41.95+575 has shown the highest degree of variability over the last $\sim$25 years. The total flux density of the source has been consistently monitored with both the VLA and MERLIN and it has been found that at both 5 GHz (C-band) and 1.6 GHz (L-band) the flux density has been decaying at a constant rate of $\sim$8.5$\%$ yr$^{-1}$ \cite{trotman96}. Based on the total flux density as measured by these instruments, the predicted flux density for 41.95+575 at epoch 1998.9 is $\sim$55 mJy, taking into account the 8.5$\%$ yr$^{-1}$ decay rate. However, the total flux density recovered from the numerous cycles of self-calibration was 43.8 $\pm$ 2.3 mJy which only represents $\sim$80$\%$ of the predicted flux density. However, we do not attribute this to a larger than expected drop in its flux density; it is more likely due to missing extended flux on the long VLBI baselines. More recent MERLIN observations (epoch 1999.1) at 5 GHz have shown no evidence for any departures from the 8.5$\%$ yr$^{-1}$ decay rate (result not presented here).

\subsubsection{Summary of results on 41.95+575}
The simplest explanation for the appearance of 41.95+575 remains that of a supernova event taking place within a high density molecular cloud which provides the high level of confinement and results in the low expansion speeds which we derive. The high ambient density of the medium surrounding the source may also account for the X-ray luminosity, since Stevens et al. (1999)\nocite{stevens99} propose the the radio source is coincident with a point source as observed by ROSAT. The higher resolution of the Chandra X-Ray Observatory will make this identification more robust once an accurate registration with a radio image has been performed. 

It is clear that an adequate supernova-related model which fully explains both the radio morphology and flux-density variability of this source has not yet been constructed. Therefore, we must also ask the question as to whether the source is really supernova-related.

\subsection{The evolution of the shell-like remnant, 43.31+592}
\subsubsection{Limits on deceleration from low-resolution maps}
\begin{figure*}
\begin{center}
\setlength{\unitlength}{1cm}
\begin{picture}(0,0)(0,0)
\put(0,0){\includegraphics{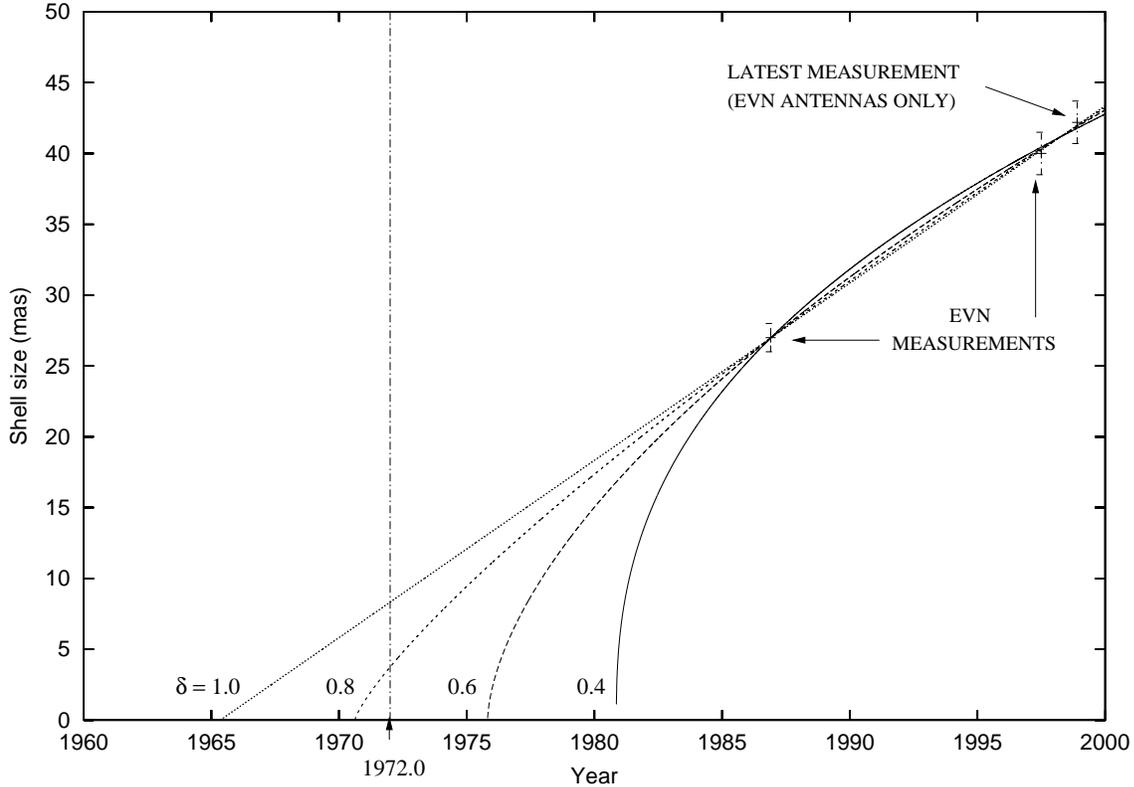}}
\end{picture}
\vspace{10.5cm}
\end{center}
\caption{Size evolution plots for the compact young supernova remnant, 43.31+592. The curves show four possibilities for the evolution with deceleration parameters from 0.4 to 1.0 (free expansion).}
\label{sizeage}
\end{figure*}

The global VLBI map of 43.31+592 represents the third epoch for which a map has been produced of sufficient resolution to image the structure of the source. In an earlier paper, Pedlar et al. (1999) described the results of the first two epochs which consisted of observations conducted by the European VLBI Network in 1986 and 1997. By taking slices at different position angles across the shell they measured angular diameters of 28.5 $\pm$ 1.0 mas and 42.1 $\pm$ 1.5 mas for epochs 1986.9 and 1997.5 respectively. These angular sizes correspond to linear sizes of 0.442 $\pm$ 0.016 pc and 0.653 $\pm$ 0.023 pc at 3.2 Mpc. Using the nomenclature of Huang et al. (1994)\nocite{huang94} we parameterise the size evolution of the remnant as,
\begin{equation}
D_{pc} = kT_{yr}^{\delta},
\end{equation}
where $D_{pc}$ is the shell diameter in parsecs, $T_{yr}$ is the age of the remnant in years, k is a constant and $\delta$ is a `deceleration parameter'. Since the date of the supernova event is unknown, there are three free parameters to which we need to fit our data. Therefore, the current data are not sufficient to constrain the parameters. In addition, the uncertainty in the age of the remnant does not allow a correction for the positive bias in the size measurement of the shell due to the finite VLBI beamsize as discussed by Marcaide et al. (1997)\nocite{marcaide97}. Figure~\ref{sizeage} shows four example size evolution curves which would be consistent with our data. However, since the remnant clearly existed in 1972, as is shown in the 8.1 GHz map of Kronberg \& Wilkinson (1975)\nocite{kronberg75}, it can be seen that the smaller values of $\delta$ are not allowed by our measurements.

A more exhaustive analysis using only two measured diameters may be conducted by deriving the relationship between the age of the remnant at the 1986.9 epoch and the deceleration parameter as
\begin{equation}
T_{1986.9} = \frac{\Delta T}{\exp{\frac{ln(D_{2}/D_{1})}{\delta}} - 1},
\label{agedelta_eqn}
\end{equation}
where $\Delta T$ is the epoch separation in years and $D_2/D_1$ is the ratio of the remnant sizes at the two epochs. Therefore, the minimum value of $\delta$ assuming a supernova event taking place before 1972 can be calculated to be 0.73 $\pm$ 0.11.

This result emphasises the importance of making images with higher spatial and hence linear resolution. Even with observations spaced by $\sim$10 years (around a third of the remnant's inferred age) it is not possible to more accurately constrain the remnant's evolutionary parameters with linear resolutions like those produced by the EVN or VLBA alone.

\subsubsection{Prospects for proper motion studies from global VLBI maps}
Figure~\ref{4331final} shows the 1998.9 epoch global VLBI map of 43.31+592 at a resolution of 4 mas. Labels A to D indicate the four most significant `knots' of radio emission which we detect. Two-dimensional Gaussian fits were performed to these knots and the results are indicated by the size and position angle of the crosses. The parameters derived by the Gaussian fits to the components are shown in table~\ref{gaussiantable}. However, we do not claim that these knots are intrinsically Gaussian in nature and this method has only been used so that an indication of the expected future evolution of the remnant can be provided. The diamond shapes in figure~\ref{4331final} indicate the positions of these knots ten years after the initial global observations, assuming a self-similar expansion speed of 10000 km s$^{-1}$ \cite{pedlar99}. However, two years should be sufficient to detect the proper motions of these components since 10000 km s$^{-1}$ corresponds to 0.66 mas yr$^{-1}$ at a distance of 3.2 Mpc. A more exhaustive analysis of the high resolution images will be performed once a second epoch map at the same resolution has been obtained.
\begin{table*}
\begin{center}
\caption{The results of 2D Gaussian fits to the brightest compact knots of radio emission in the young supernova remnant, 43.31+592. The right ascensions and declinations are quoted relative to 09$^{h}$55$^{m}$00$^{s}$ +69$^{\circ}$40'00''. It should be noted that the absolute astrometric precision of these positions has been degraded since self-calibration has been used. However, the relative precision has been retained.}
\begin{tabular}{|c|c|c|c|c|c|}\hline
Label & Right Ascension & Declination & Major Axis & Minor Axis & Position Angle\\ 
& (J2000) & (J2000) & (mas) & (mas) & (degrees)\\ \hline
A & 52.03183 $\pm$ 0.00007 & 45.4157 $\pm$ 0.0007 & 13.3 $\pm$ 1.7 & 6.5 $\pm$ 0.8 & 15.5 $\pm$ 6.6\\
B & 52.02988 $\pm$ 0.00009 & 45.4099 $\pm$ 0.0008 & 11.0 $\pm$ 1.8 & 7.0 $\pm$ 1.1 & 168 $\pm$ 14\\
C & 52.02692 $\pm$ 0.00019 & 45.4506 $\pm$ 0.0008 & 15.1 $\pm$ 2.6 & 8.8 $\pm$ 1.5 & 124 $\pm$ 12\\
D & 52.02542 $\pm$ 0.00010 & 45.4431 $\pm$ 0.0012 & 17.6 $\pm$ 2.8 & 7.2 $\pm$ 1.2 & 11.4 $\pm$ 6.4\\ \hline
\end{tabular}
\label{gaussiantable}
\end{center}
\end{table*}

\subsubsection{Comparison with other sources}
\begin{figure*}
\begin{center}
\setlength{\unitlength}{1cm}
\begin{picture}(0,0)(0,0)
\put(0,0){\includegraphics{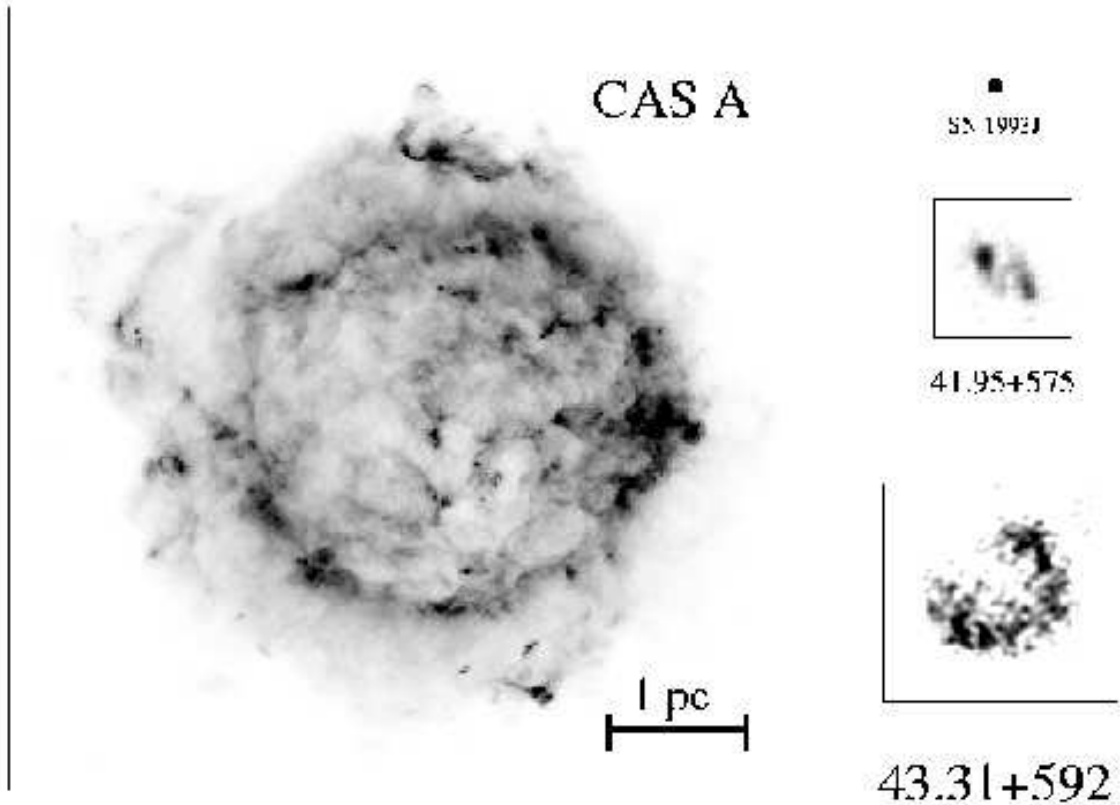}}
\end{picture}
\vspace{12cm}
\caption{Comparison of a VLA image of Cassiopeia A (Braun, Gull \& Perley 1987), the youngest supernova remnant in the Milky Way, with 43.31+592, the youngest supernova remnant in M82. Both SNR are shown on the same linear scale. Also shown on the same scale is the compact radio source, 41.95+575 and the young radio supernova SN 1993J in M81. The scale has been calculated by assuming a distance to Cas A of 2.8 kpc and a distance of 3.2 Mpc to M82 and M81. The greyscale for Cas A ranges from 0 to 0.01 Jy pixel$^{-1}$, the greyscale for 43.31+592 ranges from 0.07 to 0.2 mJy beam$^{-1}$ and the greyscale for 41.95+575 ranges from 0.1 mJy beam$^{-1}$ to 1.5 mJy beam$^{-1}$.}
\label{casa}
\end{center}
\end{figure*}

Figure~\ref{casa} shows a VLA image of the galactic supernova remnant, Cassiopeia A \cite{braun87} on the same linear scale as the compact sources in M82. Also shown schematically is the young radio supernova, SN 1993J in M81. The size of the circle corresponds to the size of SN 1993J as measured by Marcaide et al. (1997)\nocite{marcaide97} in October 1996. This figure emphasises the difference between studying galactic supernova remnants and those in M82. Cassiopeia A has an age of around 320 years \cite{ashworth80} compared with an age of $\sim$30 years for 43.31+592. It is therefore clear that studies of the young supernova remnants in M82 can fill an important gap in the studies of the evolution of supernova remnants between the very young (e.g. SN1993J) and those with ages of a few hundred years. However, any differences that are due to the M82 supernovae evolving in a starburst ISM may also become apparent.

\subsection{Other sources}
Of the remaining three sources detected by the EVN at epoch 1997.5 only 44.01+596 and 45.17+612 were reliably detected in this expermiment.
\subsubsection{44.01+596}
\begin{figure*}
\begin{center}
\setlength{\unitlength}{1cm}
\begin{picture}(0,0)(0,0)
\put(0,0){\includegraphics{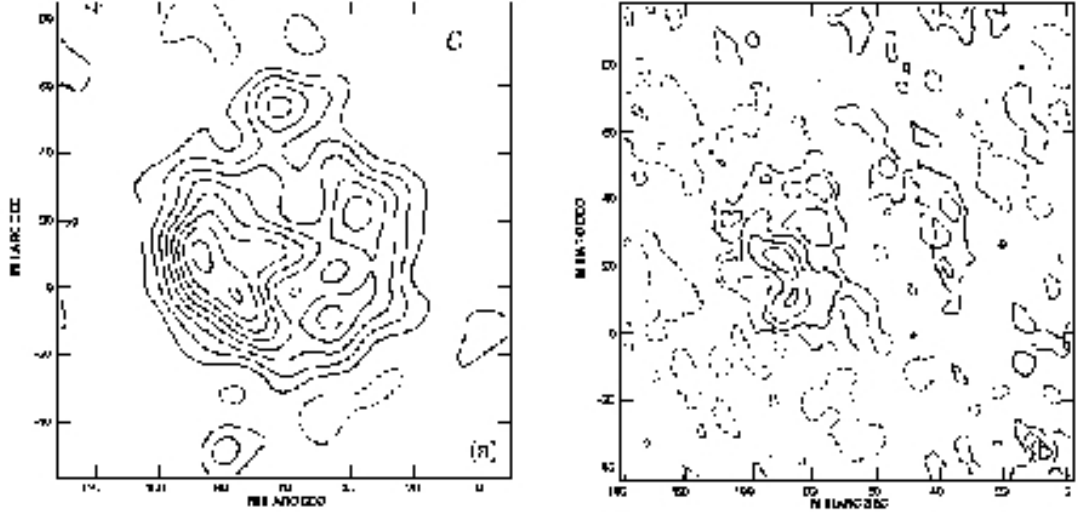}}
\end{picture}
\vspace{7cm}
\caption{Comparison between the 1997.5 epoch EVN image of 44.01+596 (15 mas resolution) with the 1998.9 epoch global VLBI image (10 mas resolution). The contour levels for both images are (-1, 1, 2,..., 9, 10) $\times$ 0.1 mJy beam$^{-1}$.}
\label{4401final}
\end{center}
\end{figure*}

Of all the compact radio sources in M82, 44.01+596 is the most likely candidate as the host of a weak active galactic nucleus \cite{wills99}. However, when imaged with the EVN at the 1997.5 epoch it showed structure more consistent with an identification as a supernova remnant. Wills et al. (1999) argued that this does not exclude the AGN possibility since Sgr A* also shows similar radio continuum structure at a similar linear scale.
Figure~\ref{4401final} shows a comparison between the 1997.5 epoch EVN image of 44.01+596 (15 mas resolution) with the 1998.9 epoch global VLBI image (10 mas resolution). The same structure that is seen in the lower resolution image is clearly present at higher resolution, although the signal to noise ratio in the 10 mas map is not sufficient that any apparent changes in morphology can be observed with any confidence. As with the other sources, more observations are required at later epochs to determine any expansion of the `shell' and hence aid its identification as a supernova remnant or weak AGN.
\section{Conclusions}

These global VLBI observations have produced the highest resolution images yet made of the young supernova remnants formed in the nuclear starburst within M82. At the full resolution of $\sim$3 mas images have been made of the brightest and what are inferred to be the youngest sources. For the compact young supernova remnant, 43.31+592, we draw the following conclusions.

(1) The measurements of the shell diameter from 15 mas resolution EVN images \cite{pedlar99} places a limit on the deceleration parameter. Since the remnant is present in the 1972 image of Kronberg \& Wilkinson (1975)\nocite{kronberg75}, a limit of $\delta >$ 0.73 $\pm$ 0.11 has been calculated. As yet, this limit does not allow for a physical analysis of the supernova model, but with future observations tighter constraints will be placed on the deceleration of this source.

(2) The higher resolution images presented in this paper will enable proper motion studies to be carried out on the compact knots of radio emission as shown in figure~\ref{4331final}. Further global VLBI observations are scheduled to take place in late 2000, two years after the initial observations. This represents the minimum time required to detect significant proper motion of the knots assuming a radial expansion speed of 10000 km s$^{-1}$.

(3) For the first time we have been able to image the supernova remnants in M82 on a similar linear scale to that which we can image the galactic supernova remnant, Cassiopeia A.

In addition, we draw the following conclusions for the compact radio source, 41.95+575.

(1) The new image of 41.95+575 reveals structures which represent deviations from its previous description as an elongated shell and we speculate that these may either represent collimated outflow as predicted by the jet-induced explosion models of Khokhlov et al. (1999) or a prominence of the type predicted by Blondin et al. (1996).

(2) We calculate an upper limit to the radial expansion rate along the major axis of 2000 km s$^{-1}$.

The young supernova remnants in M82 fill an important gap in the evolutionary timescale of supernova remnants between the very young (e.g. SN1993J) and those with ages measured in hundreds of years (e.g. those in the Milky Way and Magellanic clouds). Therefore, M82 will continue to act as an important laboratory for the study of young SNR and any differences in evolution which are due to the starburst ISM may become apparent.
\subsection*{Acknowledgements}
We acknowledge the contribution of the observatories who participate in the European VLBI Network and the VLBA. We thank Rick Perley at NRAO, Socorro for allowing us to use the VLA image of Cassiopeia A. We thank Tim O'Brien and John O'Connor for useful comments. Andrew McDonald acknowledges the receipt of a PPARC postgraduate research grant.
{}
\label{lastpage}
\newpage
\end{document}